\documentclass[aps,amsmath,amssymb,amsfonts,tightenlines,prd,10pt,twocolumn,prX,superscriptaddress,nofootinbib]{revtex4-1}
\usepackage{lipsum}
\usepackage{bm, slashed}
\usepackage{graphicx}
\usepackage{epic,eepic}
\usepackage{enumitem}

\usepackage[cm]{fullpage}

\newcounter{incre}
\newcommand{\benum}{\begin{enumerate}[label=\roman{enumi}),ref=\roman{enumi}] \setcounter{enumi}{\value{incre}}}
\newcommand{\eenum}{\setcounter{incre}{\value{enumi}}\end{enumerate}}
\newcommand{\beq}{\begin{equation}}
\newcommand{\eeq}{\end{equation}}

\newcommand{\beqa}{\begin{eqnarray}}
\newcommand{\eeqa}{\end{eqnarray}}

\begin{document}

\title{$B^\pm\to D  K^\pm $ with direct CP violation in charm}
\author{Mario Martone}
\email{mcm293@cornell.edu}
\affiliation{Laboratory for Elementary-Particle Physics, Cornell University, Ithaca, N.Y.}
\author{Jure Zupan} 
\email{jure.zupan@cern.ch} 
\def\Cincy{Department of Physics, University of Cincinnati, Cincinnati, Ohio 45221,USA}
\affiliation{\Cincy}

\date{\today}

\begin{abstract}
  We investigate the implications of direct CP violation (CPV) on the
  determination of the unitarity triangle angle $\gamma$ from $B\to
  DK$ decays. We show that $\gamma$ can still be extracted even with
  the inclusion of direct CPV in charm if $(i)$ at least one of the
  $D$ decays has negligible CP violation; and $(ii)$ data from charm
  factory at threshold are used. If approximate expressions without
  including direct CP violation in charm are used, this can result in a shift
  in $\gamma$ that is ${\mathcal O}(r_D/r_B)$. It is modest for $B\to
  DK$ but can be ${\mathcal O}(1)$ for $B\to D\pi$. We illustrate the
  size of the shift on an example of Gronau-London-Wyler method.
\end{abstract}

\maketitle
\section{Introduction}
Both LHCb \cite{LHCb:2011} and CDF \cite{CDF:2012} have measured nonzero 
CP violation in the charm sector giving a combined value for the difference of  CP asymmetries
\beq\label{LHCb}
\begin{split}
\Delta a_{CP}&=a_{CP}(K^-K^+)-a_{CP}(\pi^-\pi^+)\\
&=(-6.45\pm1.80)\times 10^{-3}.
\end{split}
\eeq
Here $a_f\equiv a_{CP}(f)$ is the time integrated CP asymmetry for $D$ decaying to a  CP eigenstate $f$
\beq
a_f\equiv\frac{\Gamma(D^0\to f)-\Gamma(\bar{D}^0\to f)}{\Gamma(D^0\to f)+\Gamma(\bar{D}^0\to f)}.
\eeq
In general $a_f$ is a sum of three contributions
\beq
a_f=a_f^{\rm dir}+a_f^m+a^i_f,
\eeq
where  $a_f^{\rm dir}$  is CP violation in the decay (or direct CP violation), $a_f^m$ CP violation from mixing and $a^i_f$ CP violation from interference of decay and mixing \cite{Grossman:2007}. Mixing effects are universal, independent of the final state, and are furthermore suppressed by the $D^0-\bar{D}^0$ mixing parameters $x,y\sim{\mathcal O}(10^{-2})$ and are in general negligible (the exact size does depend on the time interval the experiments integrate over, however). For instance, LHCb quotes 
\cite{LHCb:2011}
\beq\label{directCP}
a_{K^+K^-}-a_{\pi^+\pi^-}\approx a^{\rm dir}_K-a^{\rm dir}_\pi+(0.10\pm0.01)a_{\rm ind},
\eeq
so that $a_{\rm ind}=a^m+a^i$ can be safely neglected. 

In the present manuscript we are interested in the effect that nonzero $a_{f}^{\rm dir}$ can have on the methods for determining the CKM unitarity triangle angle, $\gamma$, \cite{Charles:2004jd}
\beq\label{gamma}
\gamma=\frac{V_{ud}V_{cb}V^*_{ub}V^*_{cd}}{V^*_{ud}V^*_{cb}V_{ub}V_{cd}}\approx(66\pm12)^\circ,
\eeq
from $B\to D^{(*)}K^{(*)}$ decays. These methods rely on the interference of $B\to D^0K\to (f)_D K$ and $B\to \bar D^0K\to (f)_D K$ amplitudes, where $f$ is a common final state to both $D^0$ and $\bar D^0$ decays \cite{Gronau:1990ra,Gronau:1991dp,Atwood:1996ci,Giri:2003ty} (see also reviews in \cite{Zupan:2011mn,Zupan:2007zz}). 
Due to the interference one can probe the relative phase of the two amplitudes, which is
related to $\gamma$. In the originally proposed methods an important
assumption was that there is no CP violation in $D$ decays\footnote{In \cite{Gronau:1990ra,Gronau:1991dp,Atwood:1996ci,Giri:2003ty} $D^0$-$\bar{D}^0$ mixing was also neglected. The effects of $x_D\neq0$ in the measurement of $\gamma$ are discussed in \cite{Silva:1998,Silva:1999,Silva:2000,Grossman:2005rp}. $D^0$-$\bar{D}^0$ mixing will be also neglected throughout the present manuscript.}. 
Below we relax this assumption and show that the effect of CP
violation in charm can be included. For a completely general treatment
additional information is needed. If this information is not included
we show how big  of an error is introduced when assuming no CPV in charm decays.

\section{General consequences of CP violation in charm}
We first focus on singly Cabibbo suppressed decays. The $D^0$ ($\bar{D}^0$) decay amplitudes $A_f$ ($\bar{A}_f$) to CP eigenstate $f$ can in general be written as 
\beqa\label{notation1}
&A_f\equiv A(D^0\to f)=A^T_fe^{i\phi_f^T}\big[1+r_fe^{i(\delta_f+\phi_f)}\big],&\\\label{notation2}
&\bar{A}_f\equiv A(\bar{D}^0\to f)=A^T_fe^{-i\phi_f^T}\big[1+r_fe^{i(\delta_f-\phi_f)}\big],&
\eeqa
where $A_f^Te^{\pm i \phi_f^T}$ is the dominant singly-Cabibbo suppressed tree amplitude. In general it can have a weak phase $\phi_f^T$, however, in the SM this is zero. Since we are primarily interested in the extraction of $\gamma$ assuming the SM, we will set $\phi_f^T=0$ in the following. The ratio $r_f$ denotes the relative magnitude of the subleading amplitude due to penguin diagrams, while $\delta_f$ and $\phi_f$ are the strong and weak phase differences, respectively. Direct CP asymmetry is then
\beq\label{adir}
a_f^{\rm dir}=-\frac{2 r_f \sin\delta_f\sin\phi_f}{1+2r_f\cos\delta_f\cos\phi_f+r^2_f}\approx-2r_f\sin\delta_f\sin\phi_f,
\eeq
where the last expression is valid for $r_f\ll1$ 
which is a good approximation in $D$ decays.

 In order to have CP violation in the SM all three generations need to participate. This means that in $D$ decays the CP violation is suppressed by the CKM elements between the third and the first two generations, giving a naive estimate  $r_f\sim \mathcal{O}\big([V_{cb}V_{ub}/V_{cs}V_{us}]\alpha_s/\pi\big)\sim10^{-4}$.
The penguin contraction matrix elements seem to be enhanced in the SM, giving an estimate $r_f\sim 10^{-3}$ and the observed size of $\Delta a_{CP}$ \cite{Brod:2011re,Brod:2012ud} (see also \cite{Pirtskhalava:2011va,Bhattacharya:2012kq,Franco:2012ck}).
The enhanced value or $r_f$ could also be due to NP \cite{Isidori:2011qw}. In this paper we are interested solely on the effect of the observed $\Delta a_{CP}$ on the extraction of $\gamma$ from $B\to DK$. We can set aside the origin of  $\Delta a_{CP}$, as our conclusion will not change, as long as the tree contributions are SM--like and do not carry weak phase.

The $B^-\to DK^-$ amplitudes are 
\begin{align}
A(B^-\to D^0K^-)&\equiv A_B,\\
A(B^-\to \bar D^0K^-)&\equiv A_B r_B e^{i(\delta_B-\gamma)},
\end{align}
where $r_B=0.099\pm0.008$, while $\delta_B=(110\pm15)^\circ$ is the strong phase \cite{Charles:2004jd}.\footnote{For simplicity of notation we focus on $B\to DK$ decay, but the results apply also to  $B\to D^*K$ and $B\to DK^*$ decays. The ratio of amplitudes and the strong phase difference are then $r_B(D^*K)=0.121^{+0.018}_{-0.019}$, $\delta_B(D^*K)=(-55^{+14}_{-16})^\circ$
and $r_B(DK^*)=0.118\pm 0.045$, $\delta_B(DK^*)=(117^{+30}_{-42})^\circ$  \cite{Charles:2004jd}.}
The amplitude for $B^- \to f_DK^-$ decay is then
\beq
\begin{split}
A(B^- \to &f_DK^-)=A_BA^T_f \big[ 1+r_B e^{i(\delta_B-\gamma)}\\
 &+ r_fe^{i(\delta_f+\phi_f)}+ r_f r_B e^{i(\delta_B-\gamma+\delta_f-\phi_f)}\big].
\end{split}
\eeq The amplitude for the CP conjugated process is obtained by flipping the signs of all the weak phases, 
$\gamma\to -\gamma, \phi_f\to -\phi_f, $ 
\beq
\begin{split}
A(B^+ \to &f_DK^+)=A_BA^T_f\big[ 1+r_B e^{i(\delta_B+\gamma)}\\
 &+ r_fe^{i(\delta_f-\phi_f)}+ r_f r_B e^{i(\delta_B+\gamma+\delta_f+\phi_f)}\big].
\end{split}
\eeq

We first discuss the errors introduced by negelecting the effect of CPV
in charm. If $r_f=0$, we have
\begin{align}
A(B^- \to &f_DK^-)=A_BA^T_f  \big[ 1+r_B e^{i(\delta_B-\gamma)}\big], \label{origAmpl}\\
 A(B^+ \to &f_DK^+)=A_BA^T_f  \big[ 1+r_B e^{i(\delta_B+\gamma)}\big]. \label{origAmpl2}
\end{align}
Now for $r_f\neq0$. Keeping just linear terms in $r_f,r_B$ we have 
\begin{align}
A(B^- \to &f_DK^-)=A_BA^T_f  \big[ 1+r_B e^{i(\delta_B-\gamma)}+r_fe^{i(\delta_f-\phi_f)}\big], \label{LinAmpl}\\
 A(B^+ \to &f_DK^+)=A_BA^T_f  \big[ 1+r_B e^{i(\delta_B+\gamma)}+r_fe^{i(\delta_f+\phi_f)}\big]. \label{LinAmpl2}
\end{align}
which can be re-written as
\begin{align}
A(B^- \to &f_DK^-)=A_BA^T_f  \big[ 1+r_B^- e^{i(\delta_B'-\gamma-\delta\gamma)}\big], \label{LinAmplexp}\\
 A(B^+ \to &f_DK^+)=A_BA^T_f  \big[ 1+r_B^+ e^{i(\delta_B'+\gamma+\delta\gamma)}\big]. \label{LinAmpl2exp}
\end{align}
Where, at $\mathcal{O}(r_f)$, $\delta\gamma=(r_f/r_B)\cos(\delta_f-\delta_B)\sin(\phi_f-\gamma)$. So the shift in $\gamma$ is ${\mathcal O}(r_f/r_B)$. This is the case since sensitivity to $\gamma$ arises at ${\mathcal O}(r_B)$ therefore all corrections need to be compared to 
the size of the smaller amplitude in \eqref{origAmpl}, \eqref{origAmpl2}. 

For each final state $f$ there is also a shift in the $\delta_B$ strong phase, $\delta_B'-\delta_B=(r_f/r_B)\sin(\delta_f-\delta_B)\cos(\phi_f-\gamma)$ as well as in the ratio of the two CKM amplitudes, which now exhibits direct CP violation $r_B^\mp-r_B=(r_f/r_B) \cos[\delta_f-\delta_B\mp(\phi_f-\gamma)]$. In order to extract the  correct value of $\gamma$ all of the above corrections need to be kept. Note that the CP asymmetry at ${\mathcal O}(r_f,r_B)$ takes a simple form
\beq\label{eq:lin:ACP}
A_{CP}(B\to f_DK)=2 r_B\sin\delta_B\sin\gamma+a_f^{\rm dir},
\eeq
so that the expressions can be easily corrected at this order by measured quantities. The CP averaged branching ratio on the other hand is
\beq\label{eq:lin:Br}
\begin{split}
Br(B\to f_DK)=A_B^2 A_f^{T2}\big[1+2 r_B&\sin\delta_B\sin\gamma\\
&+a_f^{\rm dir}\cot(\delta_f)\big],
\end{split}
\eeq
so that the correction requires the knowledge of strong phase $\delta_f$. This phase can be measured using the charm-factories as we show below. For completeness we also give  in appendix \ref{App:complete_expressions} the unexpanded expressions for branching ratios and CP asymmetries.

Next we show that in principle one can still get $\gamma$
exactly even if there is CPV in charm decays. That is we can still over-constrain the system and obtain all the unknowns directly from experiments. To show this we
first count the number of parameters. Let $n_B$ be the number of
different $B$ decays, for instance $B\to D K$, $B\to D K^*$, $B\to
D\pi$, and let $n_{CA}$ and $n_{SCS}$ be the number of Cabibbo allowed and
singly Cabibbo suppressed $D$ decays, respectively. Only SCS $D$ decays are assumed
to have nonzero CP violation, 
with both the branching ratios and CP asymmetries measured. Also, $\phi_f^T$ is assumed to be zero, which is the case in the SM. Each SCS $D$ decay depends on four parameters, A$_f^T$, $r_f,\delta_f$, and $\phi_f$ which can be reduced using the measured  branching ratio, $Br(D\to f)$, and direct CP asymmetry $a_{CP}^{\rm dir}$. We thus have two new independent parameters per SCS $D$ decay. For each $B$ decay we have three new unknowns, $A_B,r_B,\delta_B,$ and a common unknown, $\gamma$. Therefore the total number of unknowns is  
\beq
{\rm Unkn}:3n_B+1+2n_{SCS}.
\eeq
The total number of observables is the number of different $B\to f_D K$ branching ratios and CP asymmetries, which is 
\beq
{\rm Obs}:2 n_B (n_{CA}+n_{SCS}).
\eeq 
Since this grows quadratically, while the number of unknowns grows linearly, for $n$'s high enough the system can be over-constrained. For instance, for $n_B=2$, $n_{CA}=1, n_{SCS}=2$ we have 12 measurables and 11 unknowns. So at least in principle the system is solvable. We reiterate that we assumed that there is no CPV in CA decays, and as we show next this is crucial.

In order to obtain general understanding of when $\gamma$ can be extracted let us define a CP violating phase
\beq\label{alph}
\alpha_f\equiv \arg(A_f/\bar A_f).
\eeq
To first order in $r_f$, $\alpha_f=2 r_f \cos \delta_f \sin \phi_f$. For $B\to DK$ decays one then has
\beq
\begin{split}
|A(B^\mp \to f_D K^\mp)|^2&=|A_B|^2\big[|A_f|^2+r_B^2|\bar A_f|^2+\\
&2 r_B |A_f||\bar A_f|\cos(\delta_B\mp\gamma\mp\alpha_f)\big].
\end{split}
\eeq
From this we see that there is an overall shift symmetry $\gamma\to \gamma+\phi$, $\alpha_f\to \alpha_f-\phi$. This shift symmetry is not broken by measuring $a_f^{\rm dir}$. Since $\alpha_f=a_f^{\rm dir} \cot\delta_f$, a change in $\alpha_f$ can always be compensated by adjusting the unknown strong phase difference $\delta_f$.

The shift symmetry has an important implication for $\gamma$ extraction. In fact the angle $\gamma$ cannot be extracted from $B\to D K$ data alone, unless we can disentangle the $\gamma$ and $\alpha_f$ dependence, e.g. by assuming that at least one of the $D$ decays has no CP violation, $\alpha_f=0$. This is the case despite the fact that we showed above that the total number of observables can be made bigger than the number of unknowns.  The assumption $\alpha_f=0$ is valid in the SM for the Cabibbo allowed and Cabibbo doubly suppressed $D$ decays.  However, the shift symmetry is so far broken only  from the interference terms in the $B$ decay and would thus require more statistics. 


One can make further progress by using the fact that $r_f\ll1$ in any reasonable explanation of $\Delta a_{CP}$. Expanding to first order in $r_f$ one then arrives at Eqs. \eqref{eq:lin:ACP}, \eqref{eq:lin:Br}. A nice observation was made by LHCb collaboration in \cite{LHCb:2012}, where they noted that $a_f^{\rm dir}\cot(\delta_f)$ appears in the same form for $Br(B\to f_D K)$ and $Br(B\to f_D \pi)$ (but with $r_B(B\to f_D\pi)\ll1$). In the ratio $Br(B\to f_D K)/Br(B\to f_D \pi)$ thus the term $a_f^{\rm dir}\cot(\delta_f)$ cancels. The inclusion of $a_{CP}^{\rm dir}$ then gives only a marginal shift in $\gamma$ in the analysis of \cite{LHCb:2012}.

One can also use $B\to f_DK$ decays separately if the interference terms are measured  using charm factories running at $\Psi(3770)$. Then the relative phases $\alpha_f-\alpha_{f'}$ can be measured from entangled decays. For instance, for $\Psi(3770)\to D\bar D\to f_D f'_D$ decays, where both $D\to f$ and $D\to f'$ are SCS we have
\beq
\Gamma(\Psi(3770)\to f_D f'_D)\propto \alpha_f-\alpha_{f'},
\eeq
where we expanded to leading power in $r_f, r_{r'}$. If one of the two final states is Cabibbo allowed (we take it to be $f'$), then 
\beq
\Gamma(\Psi(3770)\to f_D f'_D)\propto 1- 2 \frac{\bar A_{f'}}{A_{f'}}\cos(\alpha_f).
\eeq
The ratio of DCS and CA amplitudes is small, ${\bar A_{f'}}/{A_{f'}}\sim \lambda^2\sim 0.05$, while in addition the sensitivity to $\alpha_f$ is only at quadratic order, $\cos\alpha_f=1-\alpha_f^2/2+\cdots$. Therefore the $\Psi(3770)\to D\bar D$ entangled decay with a CA decay on one side and SCS decay on the other side is not optimal for breaking the shift symmetry degeneracy. 

A better strategy is to use Dalitz plot decays, where both CA and SCS decays interfere, for instance in $D^0\to K_S\pi^+\pi^-$. In this Dalitz plot the $D^0 \to K_S \rho^0$ and $D^0\to K^{*-}\pi^+$ amplitudes interfere. We have
\beq
\begin{split}
\Gamma(D^0&\to K_S\pi^+\pi^-)\propto | A_{K_S \rho^0} BW_{\rho^0} (m_+^2,m_-^2) \\
&A_{ K^{*-}\pi^+} BW_{K^{*-}} (m_+^2,m_-^2) +\cdots|^2,
\end{split}
\eeq
where $BW_{\rho^0, K^{*-}}$ are the Breit-Wigner functions, $m_{\pm}^2=(p_{K_S}+p_{\pi^\pm})^2$, and the ellipses denote the other resonances in the Dalitz plot. For $\bar D^0$ decay we have similarly
\beq
\begin{split}
\Gamma(\bar D^0&\to K_S\pi^+\pi^-)\propto | \bar A_{K_S \rho^0} BW_{\rho^0} (m_+^2,m_-^2) \\
&\bar A_{ K^{*+}\pi^-} BW_{K^{*+}} (m_+^2,m_-^2) +\cdots|^2.
\end{split}
\eeq
Because of the known strong phase variation in the Breit-Wigner functions one can extract the phase differences $\arg(A_{K_S \rho^0}/A_{ K^{*-}\pi^+})$ and $\arg(\bar A_{K_S \rho^0}/\bar A_{ K^{*+}\pi^-})$ by measuring the corresponding interference regions in the $D^0$ and $\bar D^0$ decays.  
 Since CA decays $D^0\to K^{*-}\pi^+$  and $\bar D^0\to K^{*+}\pi^-$ do not carry a weak phase (as is the case in the SM) we also have $\arg(\bar A_{ K^{*+}\pi^-}/A_{ K^{*-}\pi^+})=0$. As a result one can extract the relative phase $\alpha_{K_S\rho^0}$ between the $D^0\to K_S\rho^0$ and $\bar D^0\to K_S \rho^0$ decay amplitudes. Once $\alpha_{K_S\rho^0}$ is measured one can use charm-factory to measure the CPV phases $\alpha_f$ for the remaining  SCS decays such as $D\to K^+ K^-, \pi^+\pi^-, \dots,$ from $\Psi(3770)\to f_D f'_D$ decays with SCS $D$ decays on both sides.

\section{Gronau-London-Wyler determination of $\gamma$}\label{sec:GLW} 

\begin{figure}[t]
\centering{
\includegraphics[width=0.35\textwidth]{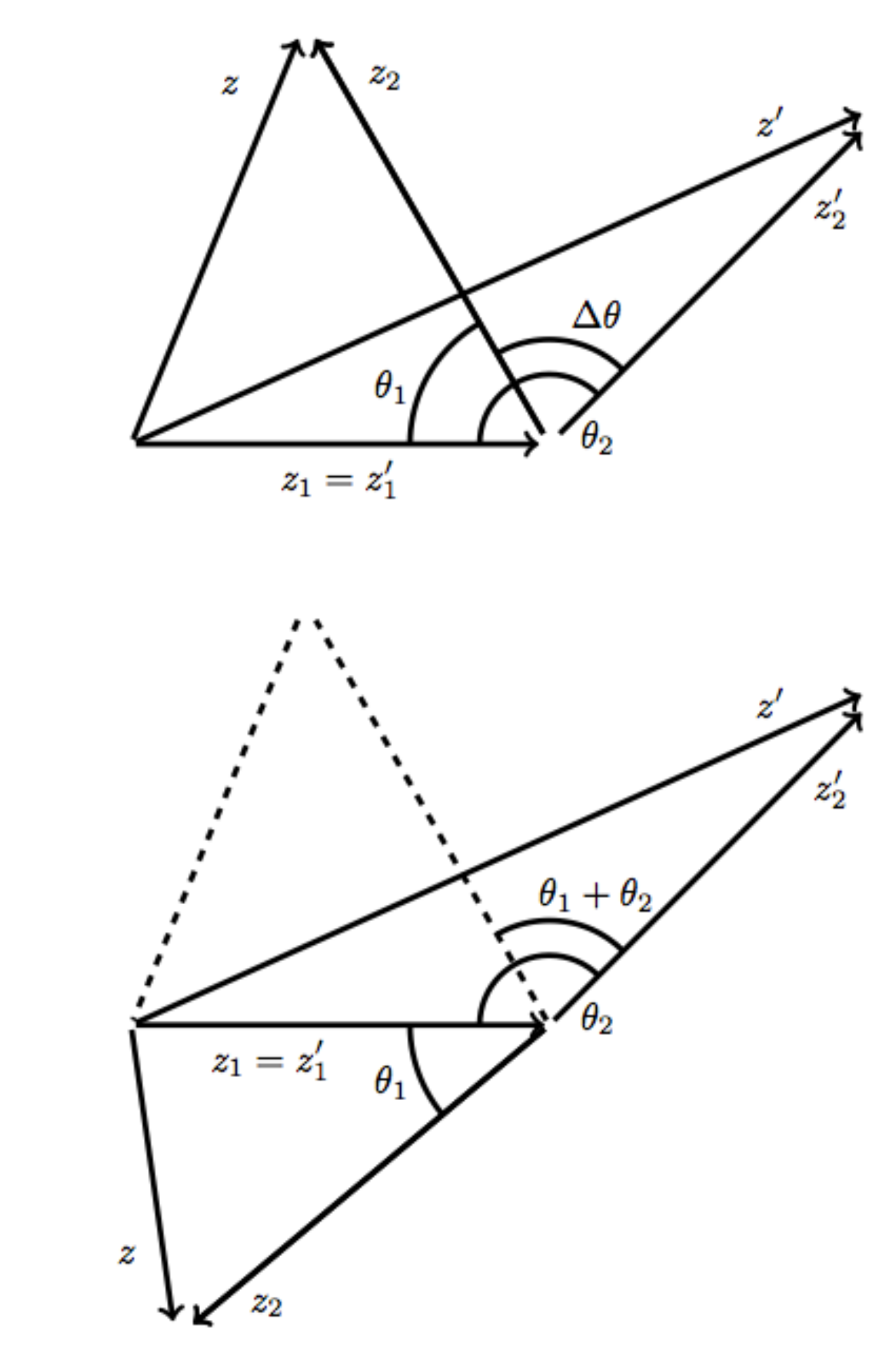}
}
\caption{\footnotesize The orientation of the triangles in GLW method with CP violation in SCS $D$ decays included. }
\label{fig:1}
\end{figure}

We now show the consequences of direct CPV in charm on determining $\gamma$ using an explicit example -- the Gronau-London-Wyler (GLW) method
\cite{Gronau:1990ra,Gronau:1991dp}. We shorten the notation following \cite{Branco:1999fs} and define
\begin{align}
\label{z1}
z_1 &\equiv \bar A_f A_B,  &z_1'&\equiv A_f A_B,\\
\label{z2}
z_2 &\equiv 	A_f A_B r_Be^{i(\delta_B+\gamma)}, &z_2'&\equiv	\bar A_f A_B r_B e^{i(\delta_B-\gamma)},
\end{align}
so that
\begin{align}
z&\equiv A(B^+ \to f_D K^+)=z_1+z_2, \label{z:triangle}\\
z' &\equiv A(B^- \to f_D K^-)=z_1'+z_2'. \label{z':triangle}
\end{align}
The magnitude $A_B$ is measured from $Br(B^-\to D^0 K^-)$, $A_B r_B$ from $Br(B^-\to \bar{D^0} K^-)$, while $|A_f|$ and $|\bar A_f|$ are obtained from measurements of the branching ratio and CP asymmetry of the $D\to f$ decays. This fixes the magnitudes $|z_{1,2}|$ and $|z_{1,2}|'$. Similarly $|z|$ and $|z'|$ are determined by measurements of $Br(B^\pm\to f_D K^\pm)$. If there is no direct CPV in charm decays then $|z_1|=|z_1'|$ and $|z_2|=|z_2'|$. This is what is assumed in the original GLW method, with the triangle constructions shown in Fig. \ref{fig:1}. The magnitudes $|z_{1,2}|$, $|z_{1,2}'|$, $|z|$ and $|z'|$  are used to construct the two triangles (with two possible orientations), while the difference of the two angles, $\theta_1={\rm arg}(z_2/z_1)$ and $\theta_2={\rm arg}(z_2'/z_1')$, determines $\gamma$,
\begin{align}
2\gamma=\theta_2 - \theta_1 = \text{arg}(z_2z_1'/z_1z_2').\label{eq:GLW:weakphase}
\end{align}

Algebraically, the difference of the two angles is given by 
\beq
2\sin^2\left( \frac{ \theta_2-\theta_1 }{ 2 } \right)=1-c_1c_2\pm\sqrt{(1-c_1^2)(1-c_2^2)},\label{sin2}
\eeq
where we shortened $c_1\equiv\cos\theta_1$ and $c_2\equiv\cos\theta_2$. The cosines are given directly in terms of the $|z_i|$ from the two triangle relations \eqref{z:triangle} and \eqref{z':triangle}. For instance $c_1 =  ({|z_1|^2 + |z_2|^2-|z|^2})/{|2 z_1z_2|}$, and similarly $c_2 =({|z_1'|^2 + |z_2'|^2-|z'|^2})/{|2 z_1'z_2'|}$.

In the derivation of \eqref{eq:GLW:weakphase} the assumption of no CPV in $D$ decays entered at two steps
\begin{itemize}
\item
{\it construction of the triangles}. The angle $\theta_1={\rm arg}(z_1/z_2)$  will no longer equal $\delta_B+\gamma$, but it will also contain the weak phase from charm decay amplitudes (and similarly for $\theta_2$). 

\item {\it overlapping of the triangles}. In general $|z_1|\ne |z_1'|$ so that the two bases of the triangles in Fig. \ref{fig:1} no longer coincide.

\end{itemize}

If there is CPV in charm decays, so that $A_f\ne \bar A_f$, then the relation \eqref{eq:GLW:weakphase} gets modified to 
\beq\label{eff11}
\begin{split}
\theta_2-\theta_1&=\text{arg} \left.\left( \frac{ z_2z_1' }{z_1z_2'}\right) \right|_{f}=2(\gamma + \alpha_f)\simeq\\
&\simeq2\big(\gamma - a^{\rm dir}_f \cot\delta_f\big),
\end{split}
\eeq
where in the last equality we have only kept terms up to ${\mathcal O}(r_f)$. The difference of the two angles, $\theta_2-\theta_1$ is thus now related to $\gamma$ through a final state dependent  phase shift $\alpha_f$ \eqref{alph}. This corrects for the first of the two points above. It also shows explicitly that there is a shift symmetry $\gamma\to \gamma+\phi$, $\alpha_f\to \alpha_f-\phi$ in the problem. So unless this shift symmetry is broken $\gamma$ cannot be determined. 

The remaining question is how $\theta_1$ and $\theta_2$ are determined. If they are determined from the original GLW construction using triangles, then there is going to be an error in the extracted value of $\gamma$. Let's denote the phases determined in this way as 
$\theta_1^{\rm dir}$ and $\theta_2^{\rm dir}$. They are given explicitly  by $c_1^{\rm dir} =  ({|z_1|^2 + |z_2|^2-|z|^2})/{|2 z_1z_2|}$ and similarly for $c_2^{\rm dir}$ with $z_i\to z_i'$. The relation between the correctly and incorrectly determined phases is obtained after some algebra to be
\beq\label{eff2}
\sin^2 \left( \frac{ \Delta \theta^{\rm dir} }{ 2 } \right)=\sin^2 \left( \frac{ \Delta \theta }{ 2 } \right) + \mathcal{C} \big( \theta_1, \theta_2 \big) \times a^{\rm dir}_f,
\eeq
where we again kept only terms up to ${\mathcal O}(r_f)$, and for shortness of notation defined $\Delta \theta^{\rm dir}\equiv \theta_2^{\rm dir}-\theta_1^{\rm dir}$ and $\Delta \theta \equiv \theta_2-\theta_1$.  The multiplicative factor in the second term is
\beq
\begin{split}\label{coeff1}
\mathcal{C} \big( \theta_1,\theta_2 \big)&=\frac{1}{4}\left( \frac{|z_1|}{|z_2|} -\frac{|z_2|}{|z_1|}\right) \tilde{\mathcal{C}} \big( \theta_1, \theta_2 \big)\simeq \frac{1}{ 4 r_B}\tilde{\mathcal{C}} \big( \theta_1, \theta_2 \big),
\end{split}
\eeq
with
\beq
\tilde{ \mathcal{C} } \big( \theta_1 , \theta_2 \big) \equiv (c_1-c_2)\Big[1+\sqrt{(1-c_1^2)(1-c_2^2)}/(c_1 c_2)\Big].
\eeq
To the order we are working this function can be evaluated using $c_i^{\rm dir}$ instead of $c_i$. 

Using GLW without taking into account CPV in charm one would conclude that $\Delta \theta^{\rm dir}$ equals $2\gamma$, a conclusion that is incorrect up to the shifts shown in \eqref{eff11} and \eqref{eff2}. To stress this let us define
\beq
\gamma^{\rm dir}\equiv  \Delta \theta^{\rm dir}/2.
\eeq
The relation between the incorrectly determined and the correct values for $\gamma$ are then from \eqref{eff11} and \eqref{eff2}
given by
\beq\label{master}
\gamma\simeq\gamma^{\rm dir} + a^{\rm dir}_f\Big[ \cot\delta_f -\frac{1}{4r_B\sin(2\gamma)} \tilde{ \mathcal{C} } (\theta_1, \theta_2) \Big],
\eeq
or numerically, using $r_B=0.1$, $\gamma=68^\circ$,
\beq
\gamma=\gamma^{\rm dir} + a^{\rm dir}_f\Big[ \cot\delta_f - 3.6 \ \tilde{ \mathcal{C} } (\theta_1, \theta_2) \Big].
\eeq

We see, that the effect vanishes in the limit of no direct CP violation, $a_f^{\rm dir}=0$. Since the sensitivity to $\gamma$ is proportional to $r_B$, part of the shift $\gamma-\gamma^{\rm dir}$ is relatively enhanced by $1/r_B$. This results in a shift that is a factor of a few times the value of $a_f^{\rm dir}$ for $B\to DK$ decays. So until the precision on $\gamma$ does not reach a level of a few percent this shift could even be ignored. This is not true for $\gamma$ extraction from $B\to D\pi$. There the ratio of the two amplitudes is in this case
\beq
r_B(D\pi)\sim \left|\frac{V_{us}V_{cd}}{V_{cs}V_{ud}}\right| r_B\sim \lambda^2 r_B\sim {\mathcal O}(0.5\%).
\eeq
Therefore the corrections due to CP violation in charm on the $\gamma$ extraction in the case of $B\to D\pi$ decays is ${\mathcal O}(1)$, since it is enhanced by an extra factor of $1/\lambda^2\approx 25$, and needs to be taken into account.

\section{Conclusions}
We have provided a strategy that can be used to correct the extraction of $\gamma$ from $B\to DK$ and $B\to D\pi$ decays for the effects of direct CP violation in charm decays. The extraction of $\gamma$ requires that there is no CP violation in Cabibbo allowed and Cabibbo suppressed $D$ decays, while the observed CP violation in singly Cabibbo allowed $D$ decays can be included using our formulas. For this the knowledge of direct CP asymmetry in $D\to f$ decays, $a_{f}^{CP}$, is needed, along with the knowledge of the relative strong phases $\delta_f$ between the interfering amplitudes in the $D$ decays. This strong phase can be obtained from high precision charm factory running at $\Psi(3770)$ in conjunction with CP violation measurements in $D$ Dalitz plot analyses, e.g. in $D\to K_S\pi^+\pi^-$. Alternatively, it can in principle be determined also in $B\to D^{(*)}K^{(*)}$ decays as well, if at least two different types of $B$ decays with several different $D\to f$ decays are considered. 
Another approach, valid to first order in the ratio of penguin and tree matrix elements in $D\to f$ SCS decays, $r_f$, is also possible \cite{LHCb:2012}. The shifts due to $a_{CP}^{\rm dir}\ne0$ in $BR(B\to f_DK)$ and $BR(B\to f_D\pi)$  are the same at ${\mathcal O}(r_f)$. The $\gamma$ extraction can thus be appropriately modified to this order to include $a_{CP}^{\rm dir}\ne 0$, if $B\to f_DK$ and $B\to f_D\pi$ decays are used simultaneously .

If CP violation in $D$ decays is ignored in the extraction of $\gamma$, the resulting shift is of the order of ${\mathcal O}(a_f^{\rm dir}/r_B)\sim$ few degrees. With increasing precision of $\gamma$ determination the inclusion of CPV in $D$ decays will therefore soon become important. For the future $\gamma$ extraction from   $B\to D\pi$ decays the shift in $\gamma$ is ${\mathcal O}(1)$ so that the inclusion of the CP violating effects will be essential from the start.
\\[-1mm]

{\bf Note added:} While this paper was being finalized Ref.  \cite{Wang:2012} appeared, which stresses the importance of including $a_f^{\rm dir}$.\\[-1mm]

{\bf Acknowledgements.} We thank Yuval Grossman for useful discussions and suggestions on the manuscript and Flip Tanedo for making Fig.1. 
The research of MM was supported in part by the NSF grant PHY-0757868. JZ was supported in part by the U.S. National Science Foundation under CAREER Grant PHY-1151392.

\appendix
\section{Complete expressions}
\label{App:complete_expressions}
In this appendix we collect the complete expressions for CP asymmetries and CP averaged branching ratios including direct CP violation in charm. The CP averaged branching ratios are
\beq
\begin{split}
\label{eq:Br:full}
{\rm Br}(&B\to f_D K)\equiv\\
\equiv&\frac{1}{2}\big(|A(B^-\to f_D K^-)|^2+|A(B^+\to f_D K^+)|^2\big)=\\
=& A_B^2 (A_f^{T})^2\Big[1+r_B^2+r_f^2+r_f^2r_B^2+\\
&+2 r_B\cos\delta_B\cos(\gamma+2\phi_f^T)+2 r_f\cos\delta_f\cos\phi_f\\
&+4r_fr_B\cos\delta_b\cos\delta_f\cos(\gamma+\phi_f+2\phi_f^T)\\
&+2 r_f r_B^2 \cos\delta_f\cos\phi_f\\
&+2 r_f^2 r_B\cos\delta_B \cos(\gamma+2\phi_f+2\phi_f^T)\Big],
\end{split}
\eeq
while the CP asymmetry is
\begin{widetext}
\beq
\begin{split}
A_{CP}(B\to f_D K)= \frac{A_B^2 (A_f^{T})^2}{{\rm Br}(B\to f_D K)}&\times
\Big[2 r_B\sin\delta_B\sin(\gamma+2\phi_f^T)+2 r_f\sin\delta_f\sin\phi_f
+4r_fr_B\sin\delta_b\cos\delta_f\sin(\gamma+\phi_f+2\phi_f^T)\\
&+2 r_f r_B^2 \sin\delta_f\sin\phi_f
+2 r_f^2 r_B\sin\delta_B \sin(\gamma+2\phi_f+2\phi_f^T)\Big],
\end{split}
\eeq
\end{widetext}
where ${\rm Br}(B\to f_D K)$ is given in \eqref{eq:Br:full}.


\begin{thebibliography}{99}

\bibitem{LHCb:2011}
R.~Aaij {\it et al.} [LHCb Collaboration],
Phys.\ Rev.\ Lett.\ {\bf 108} (2012) 111602
[arXiv:1112.0938 [hep-ex]].


\bibitem{CDF:2012} CDF collaboration, Phys. Rev. Lett. {\bf 109} (2012) 111801 [arXiv:1207.2158 [hep-ex]].

\bibitem{HFAG} Heavy Flavor Averaging Group, winter 2011, available at {\tt http://www.slac.stanford.edu/xorg/hfag}

\bibitem{Grossman:2007} Y. Grossman, A. L. Kagan, Y. Nir, Phys. Rev. D{\bf75}, 036008 (2007) [hep-ph/0609178].

\bibitem{Charles:2004jd}
J.~Charles {\it et al.} [CKMfitter Group Collaboration],
Eur.\ Phys.\ J.\ C {\bf 41} (2005) 1
[hep-ph/0406184]; Winter 2012 update.


\bibitem{Gronau:1990ra}
M.~Gronau and D.~London, Phys.\ Lett.\ B {\bf 253} (1991) 483.


\bibitem{Gronau:1991dp}
M.~Gronau and D.~Wyler,
Phys.\ Lett.\ B {\bf 265} (1991) 172.


\bibitem{Atwood:1996ci}
D.~Atwood, I.~Dunietz and A.~Soni,
Phys.\ Rev.\ Lett.\ {\bf 78} (1997) 3257
[hep-ph/9612433].


\bibitem{Giri:2003ty}
A.~Giri, Y.~Grossman, A.~Soffer and J.~Zupan,
Phys.\ Rev.\ D {\bf 68} (2003) 054018
[hep-ph/0303187].


\bibitem{Zupan:2011mn}
J.~Zupan,
arXiv:1101.0134 [hep-ph].


\bibitem{Zupan:2007zz}
J.~Zupan,
Nucl.\ Phys.\ Proc.\ Suppl.\ {\bf 170} (2007) 65.

\bibitem{Silva:1998} 
C.~C.~Meca, J.~P.~Silva,
Phys. Rev. Lett. {\bf 81} 1377-1380
[hep-ph/9807320].

\bibitem{Silva:1999} 
A.~Amorim, M.~G.~Santos, J.~P.~Silva,
Phys. Rev. D {\bf 59} (1999) 056001
[hep-ph/9807364].

\bibitem{Silva:2000} 
J.~P.~Silva, A. Soffer, 
Phys. Rev. D. {\bf 61} (2000) 112001 
[hep-ph/9912242].


\bibitem{Grossman:2005rp}
Y.~Grossman, A.~Soffer and J.~Zupan,
Phys.\ Rev.\ D {\bf 72} (2005) 031501
[hep-ph/0505270].



\bibitem{Brod:2011re}
J.~Brod, A.~L.~Kagan and J.~Zupan,
Phys.\ Rev.\ D {\bf 86} (2012) 014023
[arXiv:1111.5000 [hep-ph]].


\bibitem{Brod:2012ud}
J.~Brod, Y.~Grossman, A.~L.~Kagan and J.~Zupan,
JHEP {\bf 1210} (2012) 161
[arXiv:1203.6659 [hep-ph]].


\bibitem{Pirtskhalava:2011va}
D.~Pirtskhalava and P.~Uttayarat,
Phys.\ Lett.\ B {\bf 712} (2012) 81
[arXiv:1112.5451 [hep-ph]].


\bibitem{Bhattacharya:2012kq}
B.~Bhattacharya, M.~Gronau and J.~L.~Rosner,
arXiv:1207.0761 [hep-ph].


\bibitem{Franco:2012ck}
E.~Franco, S.~Mishima and L.~Silvestrini,
JHEP {\bf 1205} (2012) 140
[arXiv:1203.3131 [hep-ph]].


\bibitem{Isidori:2011qw}
G.~Isidori, J.~F.~Kamenik, Z.~Ligeti and G.~Perez,
Phys.\ Lett.\ B {\bf 711} (2012) 46
[arXiv:1111.4987 [hep-ph]];
Y.~Hochberg and Y.~Nir,
Phys.\ Rev.\ Lett.\ {\bf 108} (2012) 261601
[arXiv:1112.5268 [hep-ph]];
G.~F.~Giudice, G.~Isidori and P.~Paradisi,
JHEP {\bf 1204} (2012) 060
[arXiv:1201.6204 [hep-ph]];
W.~Altmannshofer, R.~Primulando, C.~-T.~Yu and F.~Yu,
JHEP {\bf 1204} (2012) 049
[arXiv:1202.2866 [hep-ph]];
T.~Feldmann, S.~Nandi and A.~Soni,
JHEP {\bf 1206} (2012) 007
[arXiv:1202.3795 [hep-ph]];
G.~Hiller, Y.~Hochberg and Y.~Nir,
Phys.\ Rev.\ D {\bf 85} (2012) 116008
[arXiv:1204.1046 [hep-ph]];
B.~Keren-Zur, P.~Lodone, M.~Nardecchia, D.~Pappadopulo, R.~Rattazzi and L.~Vecchi,
Nucl.\ Phys.\ B {\bf 867} (2013) 394
[arXiv:1205.5803 [hep-ph]];
C.~-H.~Chen, C.~-Q.~Geng and W.~Wang,
arXiv:1206.5158 [hep-ph];
C.~Delaunay, J.~F.~Kamenik, G.~Perez and L.~Randall,
arXiv:1207.0474 [hep-ph];
L.~Da Rold, C.~Delaunay, C.~Grojean and G.~Perez,
arXiv:1208.1499 [hep-ph].

\bibitem{LHCb:2012} The LHCb Collaboration, LHCb-CONF 2012-032.


\bibitem{Branco:1999fs} G. C. Branco, L. Lavoura and J. P. Silva, {\it CP Violation}, Oxford University Press (1999).

\bibitem{Wang:2012} W. Wang (2012) [hep-ph/1211.4539].

\end{thebibliography}
\end{document}